\documentclass[pre,aps,onecolumn]{revtex4}
\usepackage{amsmath,amssymb,graphicx}
\usepackage{epsfig}
\usepackage{graphicx}
\usepackage{textcomp}

\begin{document}
\title{Full absorption statistics of diffusing particles with exclusion}
\author{Baruch Meerson}
\email{meerson@mail.huji.ac.il}
\affiliation{Racah Institute of Physics, Hebrew University of
Jerusalem, Jerusalem 91904, Israel}

\begin{abstract}
Suppose that an infinite lattice gas of constant density $n_0$, whose dynamics are
described by the symmetric simple exclusion process,  is brought in contact with a spherical absorber of radius $R$.
Employing the macroscopic fluctuation theory and assuming the additivity principle, we evaluate the
probability distribution ${\mathcal P}(N)$
that $N$ particles are absorbed during a long time $T$.
The limit of $N=0$ corresponds to the survival problem, whereas $N\gg \bar{N}$
describes the opposite extreme. Here $\bar{N}=4\pi R D_0 n_0 T$ is the \emph{average} number of absorbed particles
(in three dimensions), and
$D_0$ is the gas diffusivity.
For $n_0\ll 1$ the exclusion effects are negligible, and ${\mathcal P}(N)$ can
be approximated, for not too large $N$, by the Poisson distribution with mean $\bar{N}$.  For finite $n_0$,  ${\mathcal P}(N)$ is
non-Poissonian.
We show that $-\ln{\mathcal P}(N) \simeq n_0 N^2/\bar{N}$ at  $N\gg \bar{N}$. At sufficiently large $N$ and $n_0<1/2$
the most likely density profile of the gas,
conditional on the absorption of $N$ particles, is non-monotonic in space. We also establish a close connection between this problem and
that of statistics of current in finite open systems.

\end{abstract}
\maketitle
\noindent\large \textbf{Keywords}: \normalsize non-equilibrium processes, large deviations in non-equilibrium systems, stochastic particle dynamics (theory), current fluctuations.

\tableofcontents
\nopagebreak

\section{Introduction}
\label{intro}

Statistics of large fluctuations of current in non-equilibrium steady states of diffusive lattice gases has
become  a central topic
of non-equilibrium statistical mechanics
\cite{Prahofer,Jona05,Jona06,MFTreview,Bodineau2004,DDR,Derrida07,Harris,Prolhac,Maes,Lecomte2010,Gorrisen,Akkermans,HEPG}.
The ``standard model" here involves a lattice gas between two heat baths kept at different temperatures,
or between two reservoirs of particles
at different densities. Most of the works on this subject assumed one-dimensional geometry.
It is interesting to see what new effects high dimensions can bring \cite{Akkermans,vortices}.
Here we consider a simple setting which can be studied in arbitrary dimension.
Suppose an infinite lattice gas  of density $n_0$ is brought in contact with an immobile macroscopic
(for simplicity, spherical) absorber. The gas particles are absorbed
immediately when they hit
the absorber. Here there is only one reservoir: the absorber which enforces a zero gas density in its vicinity.
This simple setting has a long history.
It was originally suggested  by
Smoluchowski \cite{S16} as a minimalistic model of diffusion-controlled binary chemical reactions (where the absorber
mimics a very large particle of the minority species). The average particle flux into the absorber
mimics the reaction rate \cite{C43,R85,OTB89,Havlin}.  Here we are interested in large fluctuations of the particle flux,
and the two main questions we ask are the following:
\begin{itemize}
\item What is the probability distribution
${\mathcal P}(N)$ that $N$ particles are absorbed during a very long time $T$? (The long-time limit is
achieved when $T$ becomes much greater
than the characteristic diffusion
time determined by the absorber radius and the gas diffusivity.)

\item What is the most probable density history of the gas,
conditional on the absorption of $N$ particles during the time $T$?
\end{itemize}

The special case of $N=0$ (all the gas particles survive until time $T$),
corresponds to  the celebrated \emph{survival} problem. This problem, and its
extensions,
have been extensively studied in the past
\cite{Donsker,ZKB83,T83,RK84,BZK84,BKZ86,BO87,Oshanin,BB03,Carlos,Havlin,KRbN,BMS13}.
Most of these studies assumed that the gas is composed of non-interacting Brownian particles in a continuous space, or non-interacting
random
walkers (RWs) on a lattice. An account of interactions between the particles (which is important, for example, in crowded environments
such as a living cell \cite{crowd}) makes the
problem much harder.  Recently, the survival problem with interactions has been addressed in  Ref. \cite{MVK}
for
diffusive lattice gases. In the hydrodynamic limit,
the coarse-grained density $n$ of these gases is governed by the diffusion equation
\begin{equation}\label{diffusion}
\partial_t n =\nabla \cdot [D(n) \nabla n],
\end{equation}
where $D(n)$ is the diffusivity. Large-scale fluctuations in these gases are described by the Langevin equation
\begin{equation}
\label{Lang}
     \partial_t n = \nabla \cdot [D(n) \nabla n] +\nabla \cdot \left[\sqrt{\sigma(n)}
     \,\text{\boldmath$\eta$} (\mathbf{x},t)\right],
\end{equation}
where $\text{\boldmath$\eta$} (\mathbf{x},t)$ is a zero-mean Gaussian noise, delta-correlated in space and in time \cite{Spohn}.
As one can see, the coarse-grained description of the fluctuations includes
an additional  transport coefficient, $\sigma(n)$. This coefficient comes from the shot noise of the microscopic model,
and it is equal to twice the mobility of the
gas \cite{Spohn}.

Here we considerably extend upon the previous work by investigating the full absorption statistics of particles
in diffusive lattice gases,  and by providing answers to the two main questions formulated above.
The long-time absorption statistics can be conveniently described by
the macroscopic fluctuation theory (MFT) of Bertini, De Sole, Gabrielli, Jona-Lasinio, and
Landim. The MFT is a variant of WKB approximation applied to Eq.~(\ref{Lang}), see
Ref. \cite{MFTreview} for a recent review. Employing the MFT, the authors of Ref. \cite{MVK}
studied the survival probability ${\mathcal P}(N=0)$ and the optimal (most likely) density history
for different lattice gases, different spatial dimensions $d$ and
different relations between the time $T$ and the characteristic diffusion time $R^2/D_0$,
where $D_0=D(n_0)$.  The simplest case turns out to be $d>2$ and $T\gg R^2/D_0$.
In this limit the leading-order results for the survival probability come from the
steady-state solution of the MFT equations which has zero flux \cite{MVK}.
Being interested in arbitrary $N$, we will assume here that the leading-order
results come from a family of stationary solutions of the MFT equations which are
parameterized by the particle flux into the absorber. A different name for the stationarity assumption is
\emph{additivity principle}. This term was coined in Ref. \cite{Bodineau2004} which
studied the statistics of current in nonequilibrium steady states (NESS) in a \emph{finite}
one-dimensional setting.

This work mostly focuses on the SSEP. In the SSEP, each particle can
hop, with an equal probability, to a neighboring lattice site if that site is unoccupied by
another particle. If it is occupied, the hop is forbidden. At the coarse-grained level, the SSEP is described
by Eq.~(\ref{Lang}) with $D(n)=D_0=\text{const}$ and $\sigma(n)=2 n_0 (1-n_0)$ \cite{Spohn,dimensions}.
For an infinite SSEP with a spherical absorber, we expect the additivity principle to hold at arbitrary $N$.

Before focusing on the SSEP we present, in Section \ref{general}, the MFT formulation
of the absorption statistics problem for an arbitrary diffusive gas at $d>2$.
Section \ref{SSEP} specifies the problem to the SSEP. A simple change of variables
maps this problem into a universal problem
of motion of an effective classical particle in a time-independent potential.
This effective mechanical problem is solved in
Section \ref{solution}, where we evaluate ${\mathcal P}(N)$ and
find the optimal density profile of the gas for arbitrary $N$ and $n_0$.
In the limit of non-interacting RWs, $-\ln{\mathcal P}(N)$ and the
corresponding optimal density profile
are determined in the Appendix.

Of special interest is the limit of $N\gg \bar{N}$ where, as we show  for the SSEP,
$-\ln{\mathcal P}(N) \simeq n_0 N^2/\bar{N}$.   As expected, this
probability density is much smaller than what is predicted by the Poisson distribution,
observed for the RWs:
 $-\ln{\mathcal P}_{\text{RW}}(N\gg 1) \simeq N \ln N$.

We also show that, for $n_0>1/2$, the optimal density profile of the SSEP,
conditional on the absorption of $N$ particles, is monotonic in space at any $N$. For $n_0<1/2$
the profile becomes non-monotonic when $N/\bar{N}$ exceeds a critical value depending on $n_0$,
see Eq.~(\ref{Nstar}).

Finally, we establish a close connection between the particle absorption statistics of the SSEP in the infinite space, considered
here at $d>2$, and the statistics of current in a finite SSEP in contact with two reservoirs at $d=1$. We show
that, when properly interpreted and rescaled, the moment generating functions of these two problems coincide.

We discuss our results and their possible extensions in Section \ref{discussion}.

\section{MFT of particle absorption: General}
\label{general}

The MFT has become a standard framework for studying large deviations
in diffusive lattice gases, see Ref. \cite{MFTreview} for a recent review.
In the MFT, the particle number density field $q(\mathbf{x},t)$ and the canonically conjugate
``momentum" density field $p(\mathbf{x},t)$ obey the Hamilton equations
\begin{eqnarray}
  \partial_t q &=& \nabla \cdot \left[D(q) \nabla q-\sigma(q) \nabla p\right], \label{d1} \\
  \partial_t p &=& - D(q) \nabla^2 p-\frac{1}{2} \,\sigma^{\prime}(q) (\nabla p)^2, \label{d2}
\end{eqnarray}
where the prime stands for the derivative with respect to the argument.   Equations~\eqref{d1} and \eqref{d2} can be written as
\begin{equation}
\partial_t q = \delta H/\delta p\,, \quad
\partial_t p = -\delta H/\delta q\,.
\end{equation}
Here
\begin{equation}
\label{Hamiltonian}
H[q(\mathbf{x},t),p(\mathbf{x},t)]= \int d\mathbf{x}\,\mathcal{H}
\end{equation}
is the Hamiltonian, and
\begin{equation}
\label{Ham}
\mathcal{H}(q,p) = -D(q) \nabla q\cdot \nabla p
+\frac{1}{2}\sigma(q)\!\left(\nabla p\right)^2
\end{equation}
is the Hamiltonian density. The spatial integration in Eq.~(\ref{Hamiltonian}), and everywhere in the following,
is performed over the whole infinite space outside the absorber. Because of the spherical symmetry of the problem,
we assume that $q$ and $p$ can only depend on the radial coordinate $r$ and time.
The boundary conditions on the absorber are \cite{Bertini,Tailleur,MR}
\begin{equation}\label{rR}
q(R,t)=p(R,t)=0.
\end{equation}
Far away from the absorber the gas is
unperturbed:
\begin{equation}\label{inf}
q(\infty, t)=n_0.
\end{equation}
A specified number $N$ of absorbed particles by time $t=T$ yields an integral constraint
on the solution \cite{DG2009b,MR,MVK}:
\begin{equation}\label{number}
\Omega_d \int_R^{\infty} dr\, r^{d-1}\,  [q(r,0)-q(r,T)] = N,
\end{equation}
where $\Omega_d =2 \pi^{d/2}/\Gamma(d/2)$ is the surface area of the $d$-dimensional unit sphere,
and $\Gamma(\dots)$ is the gamma function.

At the level of individual realizations of the stochastic process, the gas density at $t=0$ can be either deterministic or random.
In the former case (called the quenched case) one simply has
\begin{equation}\label{quenchedincond}
q(r,0)=n_0.
\end{equation}
In the latter case (called the annealed case) $q(r,0)$
is a priori unknown. As one can show \cite{DG2009b,MVK}, it obeys the following equation:
\begin{equation}\label{annealedincond}
    p(r,0)-2\int_{n_0}^{q(r,0)} dz \,\frac{D(z)}{\sigma(z)} =\lambda \theta(r-R),
\end{equation}
where $\theta(\dots)$ is the Heaviside step function, and
$\lambda$ is an a priori unknown Lagrange multiplier that is ultimately set by
Eq.~(\ref{number}).  Finally, the boundary condition for $p$ at $t=T$ is \cite{DG2009b,MR,MVK}
\begin{equation}\label{pT}
    p(r,T)=\lambda \theta(r-R).
\end{equation}

We will study the long-time particle absorption statistics in $d>2$ dimensions.
In this case,
the \emph{average} particle flux to the absorber can be found by using the \emph{stationary} solution
$n(r)$ of the diffusion equation (\ref{diffusion}). In the case of a spherical absorber, the stationary solution obeys
the equation
\begin{equation}\label{MFsteadygen}
 \frac{1}{r^{d-1}}\,\frac{d}{dr}\left[r^{d-1}\,D(n)\frac{dn}{dr} \right]=0.
\end{equation}
Solving it with the boundary conditions $n(r=R)=0$ and $n(r=\infty)=n_0$, one obtains $n=n(r)$ in implicit form:
\begin{equation}\label{implicitMF}
\frac{\int_n^{n_0} D(z) dz}{\int_0^{n_0} D(z) dz}=\left(\frac{R}{r}\right)^{d-2},\;\;\;\;\;d>2.
\end{equation}
The long-time behavior of the average number of absorbed particles $\bar{N}$ can now be found by
multiplying the particle flux to the absorber by time. The result is
\begin{equation}\label{Navgen}
\bar{N}(T)=(d-2) \Omega_d R^{d-2} T \int_0^{n_0} D(z) dz .
\end{equation}
In particular, for $D(n)=D_0 =\text{const}$ (as it happens for the non-interacting RWs, for the SSEP and
for the KMP model),  Eq.~(\ref{MFsteadygen}) becomes the Laplace's equation leading to
\begin{equation}\label{MFsteady}
    n(r)=n_0\left(1-\frac{R^{d-2}}{r^{d-2}}\right),\;\;\;\;\;d>2,
\end{equation}
and
\begin{equation}\label{NavRW}
\bar{N}(T)=(d-2) \Omega_d R^{d-2} D_0 n_0 T.
\end{equation}

We argue that, at $d>2$, fluctuations of the number of absorbed particles also come from a stationary solution,
but this time it is the stationary solution of the MFT equations (\ref{d1}) and (\ref{d2}) which account for fluctuations.
In other words, we assume
that the \emph{additivity principle}, postulated in Refs. \cite{Bodineau2004,DDR} in a finite system with two reservoirs,
holds at $d>2$ for the infinite system with one absorber. The stationary solution $q(r)$ yields, in the leading order of theory,
the optimal density profile of the
system, conditional on the number of
absorbed particles $N$. Once the steady state solutions $q(r)$ and $p(r)$ are found, we can calculate the action  which
yields $\ln {\mathcal P} (N)$
up to a pre-exponential factor:
\begin{equation}\label{actionmain}
 -\ln {\mathcal P} \simeq  \Omega_d \int_0^T dt\, \int_R^{\infty} dr\, r^{d-1}\,\left(p\partial_t q-\mathcal{H}\right)
  =\frac{1}{2}\,\Omega_d \,T \int_{R}^{\infty} dr\,r^{d-1}\,
\sigma(q)\, (\partial_r p)^2.
\end{equation}
Notice that the steady-state solutions do not obey the boundary conditions in time, Eq.~(\ref{quenchedincond}) or (\ref{annealedincond}),
and Eq.~(\ref{pT}). To accommodate these conditions, the true time-dependent solution of the problem develops two narrow boundary
layers in time,
at $t=0$ and $t=T$ that give a subleading contribution to the action, cf. Ref. \cite{MVK}.

For the spherically symmetric stationary solutions  Eqs.~(\ref{d1}) and (\ref{d2}) simplify to
\begin{eqnarray}
 \!\!\! &&r^{d-1}\left[-D(q) \frac{dq}{dr} +\sigma(q) \frac{dp}{dr}\right]= - R^{d-2} J, \label{flux}\\
\!\!\!  &&\frac{D(q)}{r^{d-1}} \,\frac{d}{dr}\left(r^{d-1}  \frac{dp}{dr}\right)+\frac{1}{2}\,\sigma^{\prime}(q) \left( \frac{dp}{dr}\right)^2= 0, \label{steadyv}
\end{eqnarray}
where we have set the negative arbitrary constant in Eq.~(\ref{flux}) to $-R^{d-2} J$, so that $J>0$.
The number of absorbed particles $N$ can be expressed via $J$ as follows:
\begin{equation}\label{NvsJ}
N=\Omega_d R^{d-2} J T.
\end{equation}
Equation~(\ref{flux}) yields
\begin{equation}\label{vviaq}
\frac{dp}{dr}=\frac{D(q)\frac{dq}{dr}-\frac{R^{d-2} J}{r^{d-1}}}{\sigma(q)}.
\end{equation}
Plugging this into Eq.~(\ref{steadyv}) we obtain
\begin{equation}
\label{steadyeqgenphys}
\frac{1}{r^{d-1}}\frac{d}{dr}\left(r^{d-1} D \frac{dq}{dr}\right)-\frac{\sigma^{\prime}D}{2\sigma}\left(\frac{dq}{dr}\right)^2
+ \frac{R^{2d-4} J^2 \sigma^{\prime}}{2 \sigma D r^{2d-2}}=0,
\end{equation}
or
\begin{equation}
\label{steadyeqgen}
\nabla^2_r q+\left(\frac{D^{\prime}}{D}-\frac{\sigma^{\prime}}{2\sigma}\right) \left(\frac{dq}{dr}\right)^2
+ \frac{R^{2d-4} J^2 \sigma^{\prime}}{2 \sigma D^2 r^{2d-2}}=0,
\end{equation}
where
$$
\nabla^2_r = \frac{1}{r^{d-1}}\,\frac{d}{dr}\left(r^{d-1}\,\frac{d}{dr} \right)
$$
is the spherically symmetric Laplace operator in $d$ dimensions.
There are two limits worth mentioning here:

\begin{enumerate}
\item In the mean-field limit the first term  in Eq.~(\ref{steadyeqgenphys}) vanishes, see Eq.~(\ref{MFsteadygen}), and
the balance of the remaining two terms yields
the average flux
\begin{equation}\label{JbarRW}
\bar{J}=\frac{r^{d-1}D(q)}{R^{d-2}} \frac{dq}{dr} = (d-2) \int_0^{n_0} D(z) dz.
\end{equation}
For the non-interacting RWs, the SSEP and the KMP model $\bar{J}=(d-2) D_0 n_0$.

\item The zero-flux limit $J=0$ is the  \emph{survival} limit: it provides a macroscopic description of the situation when not a single particle is
absorbed during the whole time $T$. In this limit
Eq.~(\ref{steadyeqgen}) reduces to Eq.~(23) of Ref. \cite{MVK}.
\end{enumerate}

Using Eq.~(\ref{vviaq}),
we can express the absorption probability distribution ${\mathcal P}$ from Eq.~(\ref{actionmain}) solely through $q(r)$:
\begin{equation}
\label{actionq}
  -\ln {\mathcal P} \simeq
\frac{1}{2}\,\Omega_d \,T \int_{R}^{\infty} dr\,r^{d-1}\,
\frac{\left[D(q)\frac{dq}{dr}-\frac{R^{d-2} J}{r^{d-1}}\right]^2}{\sigma(q)}.
\end{equation}

\section{SSEP: Mechanical analogy}
\label{SSEP}

The rest of the paper deals with the SSEP,
whereas the case of non-interacting RWs
is considered in the Appendix.
For the SSEP one has $D(q)=D_0=\text{const}$ and $\sigma(q)=2 D_0 q(1-q)$ \cite{Spohn}, and Eq.~(\ref{steadyeqgen}) becomes
\begin{equation}\label{steadySSEP}
\nabla^2_r q+\frac{2q-1}{2q(1-q)}  \left[\left(\frac{dq}{dr}\right)^2-\frac{R^{2d-4} (d-2)^2j^2}{r^{2d-2}}\right]=0,
\end{equation}
where
\begin{equation}\label{j}
    j=\frac{J}{(d-2) D_0}.
\end{equation}
In its turn, the absorption probability density (\ref{actionq}) reduces to
\begin{equation}
  \!\! -\ln {\mathcal P} \simeq\!
\frac{1}{4}\Omega_d D_0\,T \!\int_{R}^{\infty}  \frac{dr\,r^{d-1}}{q(1-q)}
\, \left[\frac{dq}{dr}-\frac{(d-2)R^{d-2} j}{r^{d-1}}\right]^2.
\label{prob2}
\end{equation}
Fortunately, the nonlinear second-order equation~(\ref{steadySSEP}) can be solved in elementary functions in any
dimension.
Let us
define new variables $\tau=(R/r)^{d-2}$ and $u=\arcsin\sqrt{q}$.
The resulting equation for $u=u(\tau)$,
\begin{equation}\label{utau}
\frac{d^2u}{d\tau^2}+\frac{2 j^2 \cos 2u}{\sin^3 2u}=0,
\end{equation}
is independent of $d$ and $R$. It describes one-dimensional motion of an effective classical particle with unit mass ($u$
is the ``coordinate" of the effective particle, $\tau$ is ``time") in the potential
$$
V(u)=2 j^2 \int^u \frac{\cos 2z \,dz}{\sin^3 2z}=-\frac{j^2}{2} \, \cot^2 2u .
$$
The energy integral is
\begin{equation}\label{energy}
\frac{1}{2}\left(\frac{du}{d\tau}\right)^2+V(u)=E=\text{const}.
\end{equation}
The original boundary conditions  $q(r=\infty)=n_0$ and $q(r=R)=0$ become $u(\tau=0)=\alpha\equiv \arcsin \sqrt{n_0}$ and $u(\tau=1)=0$,
respectively.  That is, our effective particle must depart at $\tau=0$ from the point with the coordinate $u=\alpha$,
where $0<\alpha<\pi/2$, and reach the origin $u=0$ at time $\tau=1$.

\begin{figure}
\includegraphics[width=0.4\textwidth,clip=]{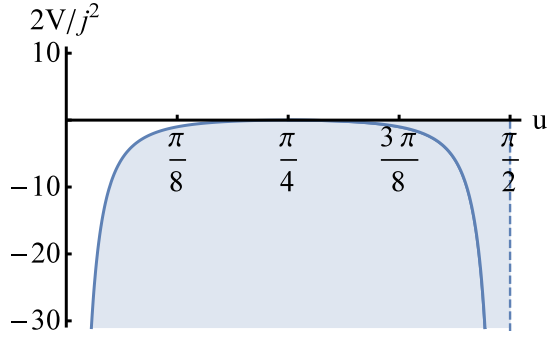}
\caption{The rescaled potential $2V(u)/j^2 = -\cot^{2} 2u$, where $0\leq u<\pi/2$.
The gray region is inaccessible for the
effective particle, see the text.}
\label{potential}
\end{figure}

Using Eq.~(\ref{energy}), we obtain
\begin{equation}\label{quadrature}
   \int\limits_{u}^{\alpha} \frac{d\xi}{\sqrt{\epsilon + \cot^{2} 2\xi}} =j \tau,
\end{equation}
where $\epsilon=2 E/j^2$ is rescaled energy of the effective particle,  to be determined from the condition
\begin{equation}\label{findconst}
   \int\limits_{0}^{\alpha} \frac{d\xi}{\sqrt{\epsilon + \cot^{2} 2\xi}} =j.
\end{equation}
The rescaled potential is  equal to $-\cot^{2} 2 u$, see Fig.~\ref{potential}.

Equations~(\ref{quadrature}) and (\ref{findconst})  assume that the effective particle only moves to the left along the $u$-axis,
so that the resulting density profile $q(r)$ is monotonic. This assumption is always correct for $\pi/4\leq \alpha<\pi/2$,
that is $1/2\leq n_0<1$. Here, in order to reach the origin, the effective particle must have a positive energy, $\epsilon>0$,
and move to the left, see Fig. \ref{potential}.

For $0\leq \alpha<\pi/4$, that is $n_0<1/2$, the gas density profile $q(r)$ is only monotonic at sufficiently small $j$,
both for positive and negative
$\epsilon$. At sufficiently large $j$ it becomes non-monotonic and develops a local maximum which is higher than $n_0$.  Here the
effective particle (with $-\cot^2 2\alpha<\epsilon<0$) first moves to the right,
is reflected from the potential barrier and then moves to the left and reaches $u=0$.  Here,
instead of Eq.~(\ref{findconst}), we need to determine $\epsilon$ from
the equation
\begin{equation}\label{findconstbounce}
   \int\limits_{0}^{\alpha} \frac{d\xi}{\sqrt{\epsilon + \cot^{2} 2\xi}} +
   2\int\limits_{\alpha}^{\alpha_r} \frac{d\xi}{\sqrt{\epsilon + \cot^{2} 2\xi}}=j,
\end{equation}
where $\alpha_r$ obeys the relation $\epsilon+\cot^2 2\alpha_r=0$, that is $2\alpha_r=\text{arccot}\,\sqrt{-\epsilon}$.
Correspondingly, $u(\tau)$, and subsequently $q(r)$, should be found from the following two equations:
\begin{eqnarray}
\label{separate}
\int_{\alpha}^u \frac{d\xi}{\sqrt{\epsilon + \cot^{2} 2\xi}}&=& j \tau, \;\;\;\;\; 0\leq \tau\leq \tau_r, \nonumber \\
\int_{u}^{\alpha_r} \frac{d\xi}{\sqrt{\epsilon + \cot^{2} 2\xi}}&=&  j(\tau-\tau_r),\;\;\;\;\; \tau_r\leq \tau\leq 1,
\label{quadraturebounce}
\end{eqnarray}
describing the effective particle moving to the right and to the left, respectively. Here
$$
\tau_r=\int_{\alpha}^{\alpha_r}  \frac{d\xi}{\sqrt{\epsilon + \cot^{2} 2\xi}}.
$$

The smaller is $n_0$, the more pronounced the non-monotonicity  of $q(r)$ becomes at large $j$. It is not surprising, therefore,
that the non-monotonicity is also present in the model of non-interacting RWs, see the Appendix.

In the variables $\tau$ and $u$, the probability distribution (\ref{prob2}) becomes
\begin{eqnarray}
 -\ln {\mathcal P} &\simeq & (d-2) \Omega_d R^{d-2} D_0 T  s (j,n_0), \nonumber \\
  s(j,n_0) &=& \int_0^1 d\tau\,\left(\frac{du}{d\tau}+\frac{j}{\sin 2u}\right)^2.
  \label{prob3}
\end{eqnarray}
The rescaled large deviation function $s(j,n_0)$ is independent of $d$, $R$ and $T$.
For concreteness, we will assume $d=3$ when presenting the formulas in dimensional (non-rescaled) form.

\section{SSEP: Solution}
\label{solution}

Before presenting the complete solution, let us consider three special cases.

\subsubsection{Mean-field limit}

This case corresponds to $j=\bar{j} = n_0 =\sin^2 \alpha$ and $N=\bar{N}$. Here,
as one can check from Eq.~(\ref{findconst}), $\epsilon=1$.
In the variables $\tau$ and $u$ the mean-field solution is
\begin{equation}\label{MFtauu}
u=\arcsin \sqrt{n_0(1-\tau)},
\end{equation}
leading to Eq.~(\ref{MFsteady}) for $d=3$. The integral in Eq.~(\ref{prob3}) vanishes, signaling the maximum
of the absorption probability distribution ${\mathcal P}$ at $N=\bar{N}$.

\subsubsection{Survival limit}
The limit of $N=0$, or $j=0$, was considered in Ref. \cite{MVK}.  Here $\epsilon$ goes to infinity, so the effective particle moves ballistically. Equation~(\ref{findconst}), with the $\cot^{2}$ term neglected, yields
$$
\int_{0}^{\alpha} \frac{d\xi}{\sqrt{\epsilon}} = \frac{\alpha}{\sqrt{\epsilon}} =j,
$$
hence
$\epsilon=\alpha^2/j^2$. Plugging this value into Eq.~(\ref{quadrature})
and again neglecting the $\cot^{2}$, we obtain  $u=\alpha(1-\tau)$. This leads to the optimal gas density profile for survival:
\begin{equation}\label{q3survival}
    q(r) = \sin^2 \left[\left(1-\frac{R}{r}\right)\,\arcsin \sqrt{n_0}\right],
\end{equation}
and to the survival probability
\begin{equation}\label{probsurvival}
-\frac{\ln {\mathcal P}}{4\pi R  D_0 T} \simeq  \int_0^1 d\tau \, \alpha^2 = \arcsin^2\sqrt{n_0},
\end{equation}
in agreement with Ref. \cite{MVK}.

\subsubsection{$N\to \infty$}

For $\pi/4\leq \alpha<\pi/2$ (that is, for $1/2\leq n_0<1$) the limit of $N\to \infty$, or $j\to \infty$,
corresponds to   $\epsilon\to +0$. Here
the effective particle (see Fig. \ref{potential}) moves to the left (if $n_0>1/2$), reaches the point $u=\pi/4$ (that is,
$q=1/2$) and spends a very long time there before finally passing through and reaching the origin. The resulting
$u(\tau)$ stays close to $\pi/4$ on most of the interval $0<\tau<1$ and has two narrow boundary layers at $\tau=0$ and $1$.
The part of the trajectory where $u\simeq \pi/4$ dominates the contribution to the probability density (\ref{prob3}),
and we obtain
\begin{equation}\label{problargej}
-\frac{\ln {\mathcal P}}{4\pi R  D_0 T } \simeq  \int_0^1 d\tau\,\frac{j^2}{\sin^2 (2\times \pi/4)} = j^2,
\end{equation}
independently of $n_0$. A dependence on $n_0$ appears when we return from $j$ to $N$, because of the relation $j=n_0 (N/\bar{N})$.

Similarly, for $0<\alpha<\pi/4$ (that is, for $n_0<1/2$), the effective particle with energy $\epsilon \to - 0$ moves to the right, reaches
the reflection point which is very close to $u=\pi/4$, spends a very long time there and then gets reflected,  moves to
the left and reaches the origin. Again,
the leading contribution to ${\mathcal P}$ is described by
Eq.~(\ref{problargej}), independently of $n_0$. Back in the physical variables we see that,  when the gas needs to pass a very
large flux to the absorber, its density
stays close to the half-filling value $n_0=1/2$ where $\sigma(q)$ is maximal, thus maximizing the fluctuation strength. The boundary
layer at $\tau=1$ becomes a boundary layer at $r=R$, whereas the boundary layer at $\tau=0$ spreads out to an infinite region $r\gg R$.

Now we determine the full absorption statistics. We first consider the case of $1/2\leq n_0<1$.

\subsection{$1/2\leq n_0<1$}

\begin{figure}
\includegraphics[width=0.4\textwidth,clip=]{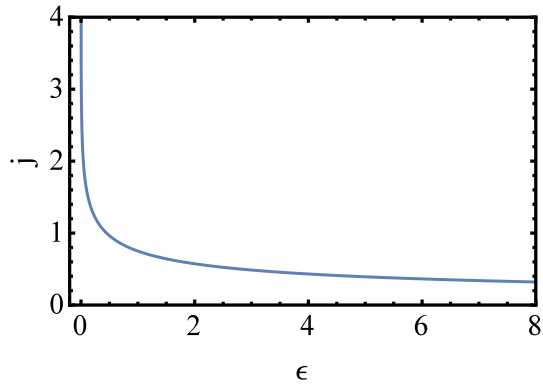}
\caption{The $\epsilon$-dependence of the function $j(\epsilon, \alpha=\pi/3)$ from Eq.~(\ref{fluxgtr}), corresponding to $n_0=3/4$.}
\label{Jvseps075}
\end{figure}

Here $\pi/4\leq \alpha<\pi/2$ and $\epsilon>0$. The effective particle can only move to the left: no reflection is possible. Evaluating the integral in Eq.~(\ref{findconst}), we obtain
\begin{equation}\label{fluxgtr}
j(\epsilon, \alpha) =
\begin{cases}
\frac{A(\epsilon, \alpha)+A(\epsilon,0)}{2\sqrt{\epsilon-1}}, & \epsilon>1,\\
\frac{B(\epsilon, \alpha)+B(\epsilon,0)}{2\sqrt{1-\epsilon}}, & 0<\epsilon<1,\\
\end{cases}
\end{equation}
where
\begin{eqnarray}
  A(\epsilon, \alpha)&=&\arctan \sqrt{\frac{\epsilon -1}{1+\epsilon
   \tan ^2 2\alpha}}, \label{A}\\
  B(\epsilon, \alpha)&=&\text{arctanh}\, \sqrt{\frac{1-\epsilon}{1+\epsilon
   \tan ^2 2\alpha}},\label{B}
\end{eqnarray}
and $\text{arctanh}\,z=\ln [\sqrt{(1+z)/(1-z)}]$.
As  $\epsilon$ varies from zero to infinity, the function $j(\epsilon, \alpha)$ monotonically decreases from plus infinity to zero.
Therefore, for $1/2\leq n_0<1$  the density profiles
can be parameterized by $\epsilon$. Note that $j(\epsilon=1,\alpha)=\sin^2 \alpha = n_0$
as expected for the mean-field solution.  The dependence of $j(\epsilon, \alpha=\pi/3)$ on $\epsilon$ is shown in Fig. \ref{Jvseps075}.

The probability density (\ref{prob3}) can be evaluated without calculating the optimal trajectory $u(\tau)$ [or, in the original
variables, the optimal density $q(r)$]. Indeed, changing the integration variable in (\ref{prob3}) from $\tau$ to $u$ and using
the energy integral (\ref{energy}) and Eq.~(\ref{fluxgtr}), we obtain
\begin{equation}\label{prob4}
-\ln {\mathcal P} \simeq  4\pi R D_0 T j(\epsilon, \alpha)
\int_0^{\alpha} du\,\frac{\left(\sqrt{\epsilon+\cot^2 2u}-\text{cosec}\, 2u\right)^2}{\sqrt{\epsilon+\cot^2 2u}}.
\end{equation}
To remind the reader, here $\pi/4\leq \alpha<\pi/2$ and  $\epsilon>0$. Evaluating the integral in Eq.~(\ref{prob4}),  we obtain
\begin{equation}\label{probgtr}
- \frac{\ln {\mathcal P}}{4\pi R D_0 T}
  \simeq j(\epsilon, \alpha) \left[(\epsilon-1) j(\epsilon, \alpha)  -\ln \sqrt{\epsilon} +
 2 \,\text{arctanh}\,\left(\frac{\sqrt{1+\epsilon \tan^2 2\alpha}+\sqrt{\epsilon} \sec 2\alpha}{1+\sqrt{\epsilon}}\right) \right].
\end{equation}
This expression is valid for all $\epsilon>0$, once we allow complex-valued functions at intermediate stages of evaluation. Equations~(\ref{fluxgtr}) and (\ref{probgtr}) determine
the probability distribution ${\mathcal P}(j)$ in a parametric  form. Its asymptotics are
\begin{equation}
\label{asympgrt}
-\frac{\ln {\mathcal P}}{4\pi R D_0T}
  \simeq
\begin{cases}
\frac{3 (j-n_0)^2}{2 n_0 (3-2n_0)}, & |j-n_0|\ll n_0, \\
\arcsin^2 \sqrt{n_0}, & j=0,\\
j^2, &j\gg n_0 .
\end{cases}
\end{equation}
As expected, the distribution is peaked at the mean-field value $j=n_0$. The distribution variance,
as represented by the Gaussian asymptotic in the first line, reaches its maximum
at $n_0=3/4$. The leading first term at $j\ll n_0$ corresponds to the survival probability \cite{MVK}. The asymptotic
at $j\to  \infty$ comes from the region where $u\simeq \pi/4$, that is $q(r)\simeq 1/2$, see Eq.~(\ref{problargej}).

\begin{figure}
\includegraphics[width=0.4\textwidth,clip=]{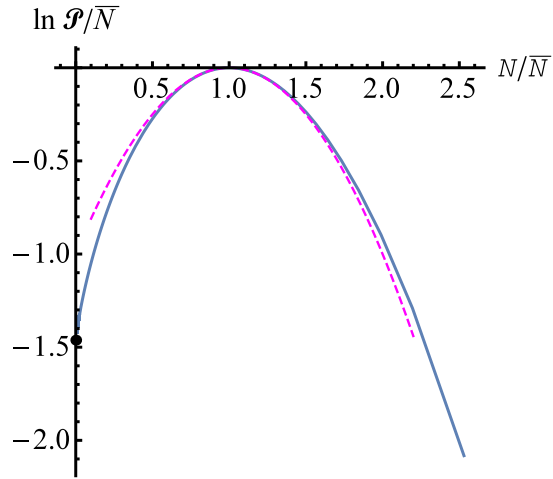}
\caption{Shown is $\ln {\mathcal P}/\bar{N}$ versus $N/\bar{N}$,
as described by Eqs. (\ref{NbyNbar}) and (\ref{SbyNbar}) for $\alpha=\pi/3$, that is $n_0=3/4$.
The dashed curve is the Gaussian asymptotics from the first line of Eq.~(\ref{asympgrt1}).
The filled point at $N=0$ shows the survival probability from the second line  Eq.~(\ref{asympgrt1}).}
\label{SSEPprob075}
\end{figure}

By virtue of the relation $j/n_0=N/\bar{N}$, Eqs.~(\ref{fluxgtr}) and (\ref{probgtr}) provides a parametric
dependence of $\ln {\mathcal P}/\bar{N}$
on $N/\bar{N}$:
\begin{eqnarray}
  &&\frac{N}{\bar{N}} = \frac{j(\epsilon,\alpha)}{\sin^2 \alpha}, \label{NbyNbar} \\
&-&\frac{\ln {\mathcal P}}{\bar{N}} =
\frac{j(\epsilon,\alpha)}{\sin^2 \alpha} \left[
(\epsilon-1) j(\epsilon, \alpha) -\ln \sqrt{\epsilon}
+ 2 \,\text{arctanh}\,\left(\frac{\sqrt{1+\epsilon \tan^2 2\alpha}
+\sqrt{\epsilon} \sec 2\alpha}{1+\sqrt{\epsilon}}\right) \right],
\label{SbyNbar}
\end{eqnarray}
whereas the asymptotics~(\ref{asympgrt}) become
\begin{equation}
\label{asympgrt1}
-\frac{\ln {\mathcal P}}{\bar{N}}
  \simeq
\begin{cases}
\frac{3 (N/\bar{N}-1)^2}{2 (3-2n_0)}, \;\;\;\;\;\;\;\;\;\;\;\;|N-\bar{N}|\ll \bar{N}, &\\
\frac{\arcsin^2 \sqrt{n_0}}{n_0},\;\;\;\;\;\;\;\;\;\;\;\;N=0, &\\
n_0 (N/\bar{N})^2, \;\;\;\;\;\;\;\;\;\;\;\; N\gg \bar{N}. &
\end{cases}
\end{equation}
Figure \ref{SSEPprob075}
depicts the probability density for $\alpha=\pi/3$, or $n_0=3/4$.

\begin{figure}
\includegraphics[width=0.4\textwidth,clip=]{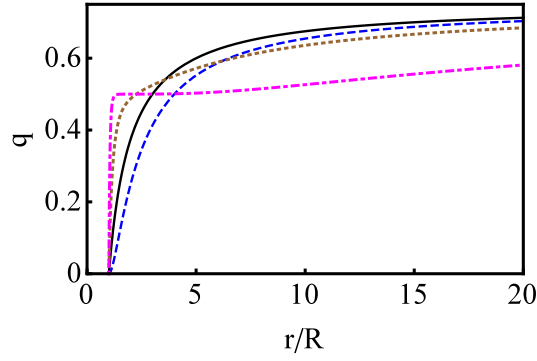}
\caption{Optimal density profiles of the SSEP, conditional on absorption of $N$ particles.
Shown is the dependence of $q$ on $r/R$
as described
by Eq.~(\ref{qgtr}) for $n_0=3/4$ and $N=\bar{N}$ (solid line), $N=0$ (dashed line),  $N=4 \bar{N}$
(dotted line) and $N=15 \bar{N}$ (dash-dotted line).
In the latter case the density stays close to $1/2$ on a large interval of distances.}
\label{q075}
\end{figure}

What is the optimal stationary density profile $q(r)$ for given $\epsilon$ and $\alpha$, that is for given $N$ and $n_0$?
Evaluating the integral in Eq.~(\ref{quadrature}), and going back to the original variables,
we obtain
\begin{equation}
\label{qgtr}
q(r,\epsilon, \alpha) =\frac{1}{2} \times
\begin{cases}
1+\sqrt{\frac{\epsilon}{\epsilon-1}}\,\sin \Phi\left(\frac{R}{r},\epsilon, \alpha\right), &\epsilon>1\\
1+\sqrt{\frac{\epsilon}{1-\epsilon}}\,\sinh \Psi\left(\frac{R}{r},\epsilon, \alpha\right), &0<\epsilon<1,
\end{cases}
\end{equation}
where
\begin{eqnarray}
  \Phi\left(\frac{R}{r},\epsilon, \alpha\right) &=& A(\epsilon, \alpha)-\left[A(\epsilon, \alpha)+A(\epsilon,0)\right]\,\frac{R}{r}, \label{Phi}\\
  \Psi\left(\frac{R}{r},\epsilon, \alpha\right)  &=& B(\epsilon, \alpha)- \left[B(\epsilon, \alpha)+B(\epsilon,0)\right]\,\frac{R}{r} \label{Psi}.
\end{eqnarray}
Equations (\ref{fluxgtr}) and (\ref{qgtr}) determine
$q(r,j,\alpha)$ in a parametric form. Examples of optimal density profiles for $n_0=3/4$ are shown in Fig. \ref{q075}.

Equations (\ref{NbyNbar}) and (\ref{SbyNbar}) simplify in the particular case $n_0=1$:
\begin{equation}\label{simple1}
-\frac{\ln {\mathcal P}}{\bar{N}} \simeq \frac{N}{\bar{N}} \left[\left(\epsilon-1\right)\frac{N}{\bar{N}} -\ln \epsilon\right].
\end{equation}
where $\epsilon$ is determined by
\begin{equation}\label{epssimple}
\frac{\arctan \sqrt{\epsilon-1}}{\sqrt{\epsilon-1}} =\frac{N}{\bar{N}},
\end{equation}
and $\bar{N}=4\pi R D_0 T \times 1=4\pi R D_0 T$.

Another  simple case is $n_0=1/2$. Here we obtain
\begin{equation}\label{simple}
-\frac{\ln {\mathcal P}}{\bar{N}} \simeq \frac{N}{2\bar{N}} \left[(\epsilon-1) \frac{N}{\bar{N}}-\ln \epsilon\right],
\end{equation}
$\epsilon$ is again determined by Eq.~(\ref{epssimple}), but now $\bar{N}=4\pi R D_0 T \times 1/2=2\pi R D_0 T$.
\vspace{0.5cm}

\subsection{$0<n_0\leq 1/2$}
Here $0 <\alpha\leq \pi/4$. At $j<j_*$, where the critical value $j_*(n_0)$ will be found shortly, the effective particle
only moves to the left, and the resulting density profile $q(r)$ is monotonic. In this regime the effective energy
$\epsilon$ can take any value between $-\cot^2 2\alpha$ and $+\infty$. Evaluating the integral
in Eq.~(\ref{findconst}), we obtain
\begin{equation}\label{fluxlss}
j(\epsilon, \alpha) =
\frac{A(\epsilon,0)-A(\epsilon, \alpha)}{2\sqrt{\epsilon-1}},\;\;\;\;\;0\leq j<j_*.
\end{equation}
where $A(\epsilon, \alpha)$ was defined in Eq.~(\ref{A}). Equation~(\ref{prob3}) again reduces to Eq.~(\ref{prob4}), but with
$0<\alpha\leq \pi/4$ and $\epsilon>-\cot^2 2\alpha$.  After some algebra,
\begin{widetext}
\begin{eqnarray}
\label{problower}
- \frac{\ln {\mathcal P}}{4\pi R D_0 T} &\simeq&  \frac{j(\epsilon,\alpha)}{4\sqrt{1-\epsilon}} \left\{\epsilon \ln \frac{2+2\sqrt{1-\epsilon}-\epsilon}{\epsilon}
+\ln \frac{2-2\sqrt{1-\epsilon}-\epsilon}{\epsilon} \right.\nonumber \\
&+& \left. 4\sqrt{1-\epsilon}\left[\ln\,\cot \alpha -\ln \left(\cot 2\alpha+\sqrt{\epsilon+\cot^2 2\alpha}\right)\right]\right. \nonumber \\
&-&\left.(\epsilon-1) \ln \frac{1-(\epsilon-1) \cos 4\alpha+\sin 4 \alpha\sqrt{(1-\epsilon) (\epsilon+\cot^2 2\alpha)}
}{\epsilon}\right\},\;\;\;\;\;0\leq j<j_*.
\end{eqnarray}
\end{widetext}
This expression is valid for $\epsilon>0$. For $-\cot^2 2\alpha<\epsilon<0$ the expression gives a complex number, and one should
take its real part.  The critical value $j=j_*$ is
achieved at $\epsilon=-\cot^2 2\alpha$:
\begin{eqnarray}
\label{criticalj}
 j_* = j(-\cot^2 2\alpha,\alpha)
 &=&(1/2) \sin 2\alpha \, \text{arctanh}\,(\sin 2\alpha) \nonumber \\
  &=& \sqrt{n_0(1-n_0)}\, \text{arctanh}\,[2\sqrt{n_0(1-n_0)}].
\end{eqnarray}
As expected, $j_*$ grows with $n_0$ and diverges at $n_0=1/2$. Indeed, at $n_0=1/2$ no reflections are possible,
and the optimal density profile $q(r)$ is monotonic for any $0\leq j<\infty$. Correspondingly, by sending
$\alpha$ to $\pi/4$
in Eqs.~(\ref{fluxlss}) and (\ref{problower}), one recovers Eq.~(\ref{simple}) for $n_0=1/2$.

\begin{figure}[ht]
\includegraphics[width=0.4\textwidth,clip=]{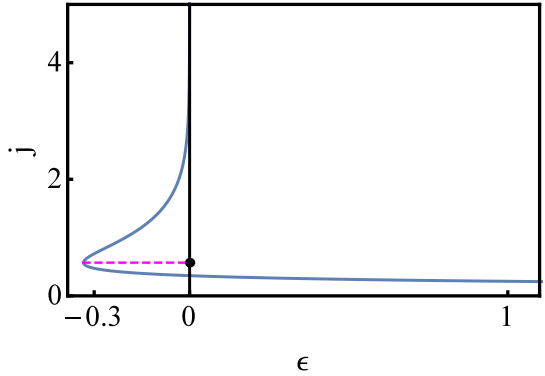}
\caption{The $\epsilon$-dependence of the function $j(\epsilon, \alpha=\pi/6)$ for $n_0=1/4$.
The lower and upper branches are described by Eqs. (\ref{fluxlss}) and (\ref{fluxlssbounce}), respectively. The filled
point indicates the critical value $j_*=0.570259\dots$ from Eq.~(\ref{criticalj}).}
\label{Jvseps025}
\end{figure}

For $n_0<1/2$ and $j>j_*$ the density profile is non-monotonic. Here $-\cot^2 2\alpha<\epsilon< 0$, and Eq.~(\ref{findconstbounce}) yields
\begin{equation}\label{fluxlssbounce}
j(\epsilon, \alpha) =
\frac{C(\epsilon,\alpha)+C(\epsilon, 0)}{2\sqrt{1-\epsilon}},\;\;\;\;\;j>j_*.
\end{equation}
where
\begin{equation}
\label{C}
C(\epsilon, \alpha)=\text{arctanh}\,\sqrt{\frac{1+\epsilon \tan^2 2\alpha}{1-\epsilon}}.
\end{equation}
A graph of $j(\epsilon, \alpha = \pi/6)$ is shown in Fig. \ref{Jvseps025}.

Now we transform from $\tau$ to $u$ in the integral Eq.~(\ref{prob3}) and account for the two parts of the trajectory:
before and after the reflection. The result is
\begin{widetext}
\begin{eqnarray}
\label{probupper}
- \frac{\ln {\mathcal P}}{4\pi R D_0 T} &\simeq& j(\epsilon,\alpha) \left\{\ln \cot \alpha -
\ln (-\epsilon)+ \frac{1}{2} \ln \left(\cot 2 \alpha+\sqrt{\epsilon+\cot^2 2\alpha}\right)  \right.\nonumber\\
  &+& \frac{1}{4}\ln\left[\epsilon+2\cot 2\alpha \left(\cot 2\alpha+\sqrt{\epsilon+\cot^2 2\alpha}\right)\right]
  -\frac{1}{4} \sqrt{1-\epsilon} \, \left\{ \vphantom{\frac{A}{B}} \ln \left(2+2\sqrt{1-\epsilon}-\epsilon\right)\right.\nonumber\\
   &-& \left.\left. \ln \left[\epsilon-\cot 2\alpha \left((\epsilon-2) \cot 2\alpha+2 \sqrt{(1-
  \epsilon)(\epsilon+\cot^2 2\alpha)}\right)\right]-\frac{1}{2}\ln (\sin 2\alpha) \right\}\right\}, \;\;\;\;\;j>j_*.
\end{eqnarray}
\end{widetext}

\begin{figure}
\includegraphics[width=0.4\textwidth,clip=]{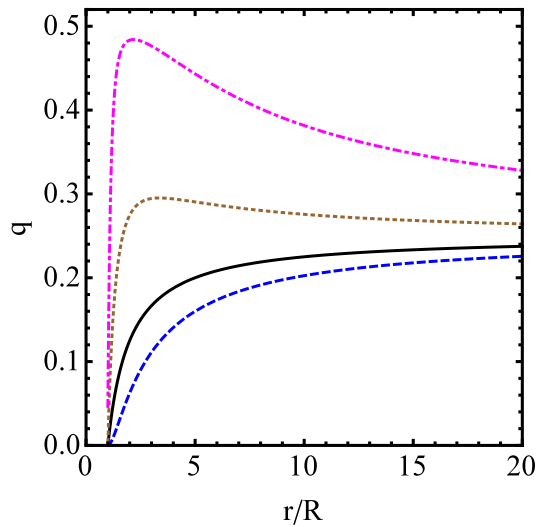}
\caption{Optimal density profiles of the SSEP, conditional on absorption of $N$ particles. Shown is the dependence of $q$ on $r/R$
for $n_0=1/4$ and $N=\bar{N}$ (solid line), $N=0$ (dashed line),  $N=4 \bar{N}$ (dotted line) and $N=15 \bar{N}$ (dash-dotted line).
At still large $N$ (not shown) the density stays close to $1/2$ on a large interval of distances.}
\label{q025}
\end{figure}

Equations~(\ref{fluxlss}), (\ref{problower}), (\ref{fluxlssbounce}) and (\ref{probupper}) determine
the probability distribution ${\mathcal P}$ versus $j$ for $n_0\leq 1/2$ and arbitrary $0\leq j<\infty$.
To go over from $j$ to $N$, as in Eqs.~(\ref{NbyNbar}) and (\ref{SbyNbar}),
one should use the relation $N=4\pi R j D_0T$ [see Eq. (\ref{NvsJ}) for $d=3$].
The critical value $j=j_*$ determines the critical number $N=N_*$ of absorbed particles
\begin{equation}\label{Nstar}
 N_* = 4\pi R D_0 T \,\text{arctanh}\,[2 n_0 (1-n_0)]
   = \bar{N} \,\frac{\text{arctanh}\,[2 n_0 (1-n_0)]}{n_0},\;\;\;\;\;n_0<1/2.
\end{equation}
At $N>N_*$ the density
profile $q(r)$ is non-monotonic.  The asymptotics~(\ref{asympgrt}) and (\ref{asympgrt1}), obtained for $1/2\leq n_0<1$,
hold for $n_0<1/2$ as well.

One can also calculate the optimal density profiles at different $N/\bar{N}$ from Eq.~(\ref{quadrature}) and (\ref{quadraturebounce}), for
$N<N_*$ and $N>N_*$, respectively.  The resulting formulas are quite cumbersome; it is much simpler to solve Eq.~(\ref{utau}) numerically by a
shooting method. Several examples of the optimal density profiles are shown in Fig. \ref{q025}.

\section{Universality of the absorption statistics}
\label{universality}

In most of the paper we have dealt with the absorption probability distribution ${\mathcal P}$, and the rescaled large deviation function $s(j,n_0)$, see Eq. (\ref{prob3}). An alternative
description of the absorption statistics is in terms of
a rescaled moment generating function $\mu(\lambda,n_0)$, the Taylor expansion of which at $\lambda=0$ yields the
distribution cumulants, see e.g. Ref. \cite{Bodineau2004}. A natural definition of $\mu$, for $d=3$, is the following:
\begin{equation}\label{mu}
\mu(\lambda,n_0) = \lim_{T\to \infty}\frac{\ln \langle e^{\lambda N(T)} \rangle}{4 \pi R D_0T} =
\lim_{T\to \infty}\frac{\ln \langle e^{4 \pi \lambda j R D_0 T} \rangle}{4 \pi R D_0T},
\end{equation}
where the averaging is with the distribution ${\mathcal P}$. A saddle-point calculation yields
\begin{equation}\label{mu2}
\mu(\lambda,n_0)=\max_{j} (\lambda j -s),
\end{equation}
where $s=s(j,n_0)$ is the rescaled action  defined in Eq.~(\ref{prob3}). The final result for $\mu(\lambda,n_0)$ is
\begin{equation}\label{mu3}
\mu(\lambda,n_0)=\,\text{arcsinh}^2 \sqrt{\omega},
\end{equation}
where
\begin{equation}
\label{omega}
\omega=n_0(e^{\lambda}-1).
\end{equation}
Let us compare this result with the rescaled moment generating function $\mu_1(\lambda,\rho_a, \rho_b)$ that describes
the statistics of current in a
one-dimensional SSEP: a chain of $L\gg 1$ lattice sites, connected at its two ends to two point-like reservoirs
at densities
$\rho_a$ and $\rho_b$. The generating function $\mu_1$ is defined
as follows:
\begin{equation}\label{mu1def}
\mu_1(\lambda,\rho_a, \rho_b) = \lim_{T\to \infty}\frac{L \ln \langle e^{\lambda J(T)} \rangle}{D_0 T}=
\lim_{T\to \infty}\frac{L \ln \langle e^{\lambda j_1 T/L} \rangle}{D_0 T}.
\end{equation}
It was calculated in Refs. \cite{Bodineau2004,DDR}, and the result is
\begin{equation}\label{mu1}
\mu_1(\lambda,\rho_a, \rho_b) =\,\text{arcsinh}^2 \sqrt{\omega_1},
\end{equation}
where
\begin{equation}\label{omega1}
\omega_1 = \rho_a (e^{\lambda}-1)+\rho_b (e^{-\lambda}-1)-\rho_a\rho_b (e^{\lambda}-1) (e^{-\lambda}-1).
\end{equation}
As we can see, $\mu$ and $\mu_1$ coincide exactly if we identify the spherical absorber with one reservoir
[and set $\rho_b=0$ in Eq.~(\ref{omega1})] and identify infinity with the other reservoir [and set $\rho_a =n_0$]. This coincidence
is unexpected because the two settings, the finite and infinite, are different. Moreover, $\mu_1$ was
obtained for $d=1$, whereas our
$\mu$ does not apply for $d<3$, where the optimal density profile is time-dependent \cite{MVK}.
The coincidence of $\mu$ and $\mu_1$ is even more interesting in view of the fact that the generating
function $\mu_1$ also describes the full counting statistics of free fermions transmitted through multichannel disordered
conductors \cite{Beenakker,Blanter,Levitov}.

The formal reason why $\mu=\mu_1$ becomes clear in the mechanical analogy of Sec. \ref{SSEP}.
Indeed, repeating our derivation for the finite one-dimensional setting we again arrive at Eq.~(\ref{utau}), except that now $\tau=x/L$, whereas the flux
$j$ is replaced by the (minus) rescaled current $j_1$. For $\ln {\mathcal P}(T)$ we obtain
\begin{eqnarray}
 -\ln {\mathcal P} &\simeq & \frac{D_0 T}{L} \,s (j_1,n_0), \nonumber \\
  s(j_1,n_0) &=& \int_0^1 d\tau\,\left(\frac{du}{d\tau}+\frac{j_1}{\sin 2u}\right)^2,
  \label{prob1d}
\end{eqnarray}
which coincides with Eq.~(\ref{prob3}) up to rescaling. In particular, the optimal density profiles in the two settings
coincide up to the coordinate transformation $x/L \leftrightarrow (R/r)^{d-2}$.

\section{Discussion}
\label{discussion}

Assuming the additivity principle, we evaluated the long-time probability distribution  ${\mathcal P}(N)$ of absorption of the SSEP by
a spherical absorber in an
infinite space.

In the low-density limit, the exclusion effects can be neglected (see the Appendix)  and, for not too large $N$,
${\mathcal P}(N)$ can be approximately described  by the Poisson distribution with mean $\bar{N}$.
For finite $n_0$,  ${\mathcal P}(N)$ is strongly non-Poissonian. In particular,  $-\ln{\mathcal P}(N) \simeq n_0 N^2/\bar{N}$ at
$N\gg \bar{N}$. This probability is much smaller than what predicted by the Poisson distribution,
$-\ln{\mathcal P}_{\text{RW}}(N\gg \bar{N}) \simeq N \ln N$, but much greater than the one predicted by the $-\ln{\mathcal P}(N) \sim N^3$
asymptotic, expected for the same setting at $d=1$ \cite{DG2009a,DG2009b,MS2014}.

At  $n_0<1/2$, and $N$ larger than the critical value $N_*$ from Eq.~(\ref{Nstar}), the most probable density profile of the gas,
conditional on the absorption of $N$ particles, is non-monotonic in space. This feature also holds for the non-interacting RWs.

An important finding of this work is a close connection between this problem and that of statistics of current
in finite systems driven by the boundaries. It was realized recently \cite{Akkermans}
that the function $\mu_1(\lambda,\rho_a, \rho_b)$ from Eqs.~(\ref{mu1}) and (\ref{omega1}), originally derived for
$d=1$ \cite{Bodineau2004,DDR},
also describes the (properly rescaled) moment generating function for $d=2$, when the two point-like reservoirs are kept
at a large but finite distance
from each other. This happens for a broad class of lattices, and different geometries. Our results extends the universality
of the generating function $\mu_1$ to a (spherically symmetric) infinite setting in any dimension greater than $2$. An immediate further extension is to
replace our spherical absorber by a spherical reservoir which enforces a non-zero density $0<\rho_b<1$,
different from the density $\rho_a=n_0$ at infinity.  The full statistics of particle absorption/emission in this setting should be
still describable in terms of the function $\mu_1$ from Eqs.~(\ref{mu1})
and (\ref{omega1}). Notice that there is complete symmetry -- both at the level of the probability distribution ${\mathcal P}$
and the optimal density profile -- with respect to the interchange of $\rho_a$ and $\rho_b$, $(R/r)^{d-2}$ and $1-(R/r)^{d-2}$,
and $N$ and $-N$. This symmetry has the form of a fluctuation theorem.

In the special case of $\rho_b\neq 0$ and $\rho_a=n_0=0$
one obtains the (long-time asymptotic of) full statistics of particle \emph{emission} from a spherical emitter into vacuum. This setting
is very similar to that of Ref. \cite{SEP_source}, except that the emitter in Ref. \cite{SEP_source} was point-like.

By virtue of the universality, we can say that the probability distribution of observing
a very large rescaled current $j_1$ in the finite one-dimensional open system \cite{Bodineau2004,DDR} should behave
as $-\ln {\mathcal P}(j_1) \sim j_1^2$. This leading contribution comes from the flat part of the density profile at the half-filling density.
To accommodate the boundary conditions at the reservoirs, the optimal density profile must develop narrow boundary
layers  at $x=0$ and $x=L$.

An interesting unresolved question is whether the universality holds if the absorber is not spherically symmetric.
It would be also interesting to study the full absorption statistics (or rather
the full energy transfer statistics) for the KMP model \cite{KMP}. For the KMP model with periodic boundaries the additivity principle breaks down at sufficiently large currents \cite{Jona05}. The breakdown mechanism boils down to the fact that the optimal density history becomes time-dependent and exhibits a traveling wave pattern \cite{Bodineau2005,BD2008,Jona06,HEPG,ZM}. It has been found recently that,
for the KMP model and a class of other models, optimal density histories of the traveling wave type are quite common, and appear for different boundary conditions \cite{MS2013}. It would be interesting to see whether the additivity principle is violated, at sufficiently
large $N$, in the settings considered in the present work.

\medskip

\section*{Acknowledgments}
I thank T. Bodineau, P.L. Krapivsky, V. Lecomte, P.V. Sasorov, O. Shpielberg and A. Vilenkin for useful discussions. This research was supported by grant No.\ 2012145 from the
United States--Israel Binational Science Foundation (BSF).

\appendix

\section*{Appendix. Full absorption statistics of non-interacting random walkers}
\label{RWs}

For non-interacting random walkers (RWs), the problem of full absorption statistics can be solved
exactly, by calculating the relevant ``microscopic" single-particle probabilities, multiplying them and extracting the long-time
asymptotics.  A simple one-dimensional example of such a calculation is presented in Ref. \cite{MR}, where all the RWs were initially released
at a single point.  See also the Appendix of Ref.
\cite{MVK} for a three-dimensional calculation in the particular case of $N=0$. Here we will directly probe the long-time regime
by employing the MFT formalism for $d=3$. At small $n_0$ (and not too large $N$, see below), we can replace the rescaled
potential $-\cot^{2} 2 \xi$  by $-(2\xi)^{-2}$.
For sufficiently small $j$, there are not reflections, and
Eqs. (\ref{quadrature}) and (\ref{findconst}) become
\begin{equation}\label{quadratureRW}
   \int\limits_{\sqrt{q}}^{\sqrt{n_0}} \frac{d\xi}{\sqrt{\epsilon + 1/(4\xi^2)}} =j \tau,
\end{equation}
and
\begin{equation}\label{findconstRW}
   \int\limits_{0}^{\sqrt{n_0}} \frac{d\xi}{\sqrt{\epsilon + 1/(4\xi^2)}} =j,
\end{equation}
respectively. The same equations are obtained if, instead of the SSEP, one considers  from the start the RWs
with $D=D_0$ and $\sigma(q)=2D_0 q$ \cite{Spohn}.
Equation~(\ref{findconstRW}) yields
\begin{equation}\label{branch1}
j=\frac{\sqrt{4 n_0 \epsilon +1}-1}{2 \epsilon }.
\end{equation}
The maximum value of $j$ in this case, $j_*=2 n_0$,  is achieved at $\epsilon=-1/(4 n_0)$.
Therefore, monotonic density profiles are obtained at $j<2 n_0$, that is at $N<2 \bar{N}$.

\begin{figure}
\includegraphics[width=0.4\textwidth,clip=]{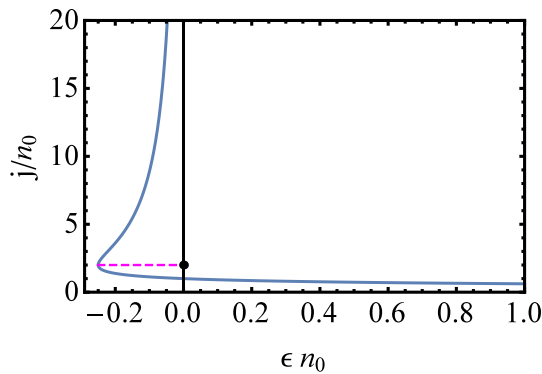}
\caption{The dependence
of $j/n_0$ on $\epsilon n_0$ for the non-interacting RWs. The lower and upper branches are
described by Eqs. (\ref{branch1}) and (\ref{branch2}), respectively.
The filled point indicates the critical value $j_*=2 n_0$.}
\label{branches}
\end{figure}

\begin{figure}
\includegraphics[width=0.4\textwidth,clip=]{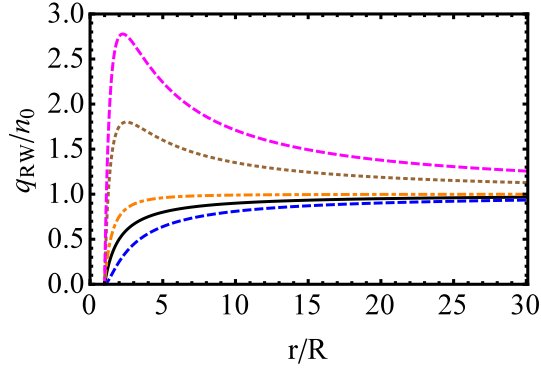}
\caption{Optimal density profiles of a gas of non-interacting RWs,
conditional on absorption of $N$ particles. Shown is $q_{\text{RW}}(r)/n_0$ versus $r/R$
as described
by Eq.~(\ref{qRW}) for $N=\bar{N}$ (solid line), $N=0$ (lower dashed line),
$N=2\bar{N}$ (dash-dotted line),  $N=6\bar{N}$ (dotted line), and $N=10 \bar{N}$ (upper dashed line).}
\label{RWdensity}
\end{figure}

Equation (\ref{branch1}) can be solved for $\epsilon$:
\begin{equation}\label{epsilonRW}
\epsilon=\frac{n_0-j}{j^2}.
\end{equation}
Evaluating the integral in (\ref{quadratureRW}) with this value of $\epsilon$ and going back to
to the original variable $r$, we obtain the optimal density profile
\begin{equation}\label{qRW}
 q_{\text{RW}}(r)= n_0\left(1-\frac{R}{r}\right) \left(1-\frac{R}{r} + \frac{jR}{n_0 r} \right)
   = n_0\left(1-\frac{R}{r}\right) \left(1-\frac{R}{r} + \frac{NR}{\bar{N}r} \right)
\end{equation}

For $j>2n_0$, or $N>2 \bar{N}$, there is reflection, and Eq.~(\ref{findconstbounce}) yields
\begin{equation}\label{branch2}
j=-\frac{\sqrt{4 n_0 \epsilon +1}+1}{2 \epsilon }.
\end{equation}
As a result, the function $j(\epsilon)$ has two branches: the lower and the upper, described
by Eqs.~(\ref{branch1}) and (\ref{branch2}), respectively, see Fig. \ref{branches}.
The inverse function $\epsilon=\epsilon(j)$,
however, is single-valued, and Eq.~(\ref{epsilonRW}), as well as Eq.~(\ref{qRW}),  remain
valid for any $0\leq j<\infty $ and, therefore, for any $0\leq N<\infty$.

Figure \ref{RWdensity}
shows the optimal density profiles for different values of $N/\bar{N}$.
For $j=n_0$ (that is, $N=\bar{N}$; this corresponds to $\epsilon=0$) it is the mean-field profile (\ref{MFsteady}) with $d=3$.
The case of $j=0$ corresponds to the survival limit $N=0$, where $q(r)=n_0(1-R/r)^2$ in
agreement with \cite{MVK}. Also shown are the cases of $N=2\bar{N}$, where
$q(r)=n_0(1-R^2/r^2)$, and two cases with still larger $N$, where the density profiles are non-monotonic. At large $N$
the density profiles become
very steep close to the absorber.

$\ln {\mathcal P}$ can be calculated from Eq.~(\ref{prob3}) where we replace $\cot 2u$ by $(2u)^{-1}$.
Straightforward calculations yield
\begin{equation}
  -\frac{\ln {\mathcal P}_{\text{RW}}}{4 \pi R D_0 T}  \simeq
j \ln \frac{j}{n_0}-j+n_0 , \label{probRW}
\end{equation}
for both non-reflecting and reflecting trajectories of the effective particle, that
is for all $N$.
Using the relation $j/n_0=N/\bar{N}$, Eq.~(\ref{probRW}) can be rewritten as
\begin{equation}
\label{probRW2}
   -\ln {\mathcal P}_{\text{RW}}  \simeq   N \ln \frac{N}{\bar{N}}-N+\bar{N},
\end{equation}
which coincides with the $N\gg 1$ limit of the Poisson distribution with mean $\bar{N}$:
$$
{\mathcal P}_{\text{Poisson}} (N,\bar{N})=\frac{\bar{N}^N e^{-\bar{N}}}{N!},
$$
In the survival limit $N=0$ Eq.~(\ref{probRW2}) yields $-\ln {\mathcal P}_{\text{RW}} \simeq \bar{N}$ in agreement with previous studies
on the survival of random walk in 3d,
see Refs. \cite{BB03,MVK} and references therein.
Close to its peak at $N=\bar{N}$, ${\mathcal P}_{\text{RW}}$ is approximately Gaussian with variance $\bar{N}$:
\begin{equation}\label{RWgauss}
-\ln {\mathcal P}_{\text{RW}}\simeq \frac{\left(N-\bar{N}\right)^2}{2\bar{N}}.
\end{equation}
As expected, the Gaussian asymptotic coincides with that obtained for the SSEP in the limit of $n_0\to 0$,
see the first line of Eq.~(\ref{asympgrt1}).

The large-$N$ asymptotic,  $-\ln {\mathcal P}_{\text{RW}}\sim N \ln N$ holds for the RWs. For the SSEP with a
small but fixed $n_0$,
this asymptotic breaks down at sufficiently large $N$, and crosses over to the large-$N$ asymptotic for the SSEP,
$-\ln {\mathcal P} \sim N^2$, see the third line
of Eq.~(\ref{asympgrt1}). The crossover occurs when the maximum of the optimal density profile becomes comparable
to $1$. As one can check,
this happens when $j$ becomes comparable with $1$, or $N$ becomes comparable with $\bar{N}/n_0$.

Finally, we note that ${\mathcal P}_{\text{RW}}$  can be approximated by the Poisson distribution only in the
leading order of theory.
This fact becomes clear already in the particular case of $N=0$, where the important subleading term,
calculated in Ref. \cite{MVK} is non-universal, as it
depends on the initial conditions.


\begin{thebibliography} {99}
\bibitem{dimensions} As it is customary in theory of lattice gases, the lattice constant 
is set to unity in all our formulas.
\end{thebibliography}
\end{document}